\documentclass[floatfix,superscriptaddress,twocolumn,showpacs,prl,amsmath,amssymb]{revtex4-1} 
\usepackage{graphicx} % Include figure files
\usepackage{color}
\usepackage{amsmath}
\usepackage{amsfonts}
\usepackage{amssymb}
\usepackage{bm}
\usepackage{txfonts}
\newcommand{\cro}{Ca$_{2}$RuO$_4$}

\newcommand{\CRO}{Ca$_{3}$Ru$_{2}$O$_7$}

\newcommand{\crto}{Ca$_{3}$(Ru$_{1-x}$Ti$_x$)$_{2}$O$_7$}

\newcommand{\Ta}{\textit{T}$_{\mathrm{N1}}$}
\newcommand{\Tb}{\textit{T}$_{\mathrm{N2}}$}

\newcommand{\Ts}{\textit{T}$_{\mathrm{MIT}}$}
\newcommand{\Td}{\textit{T}$_{\mathrm{DM}}$}
\newcommand{\rev}[1]{#1}

\begin{document}

\title{\rev{\textit{In situ} control of diamagnetism by electric current in \crto }}

\author{Chanchal~Sow}
\affiliation{Department of Physics, Graduate School of Science, 
Kyoto University, Kyoto 606-8502, Japan}
\author{Ryo~Numasaki}
\affiliation{Department of Physics, Graduate School of Science, 
Kyoto University, Kyoto 606-8502, Japan}
\author{Giordano~Mattoni}
\affiliation{Department of Physics, Graduate School of Science, 
Kyoto University, Kyoto 606-8502, Japan}
\author{Shingo~Yonezawa}
\affiliation{Department of Physics, Graduate School of Science, 
Kyoto University, Kyoto 606-8502, Japan}
\author{Naoki~Kikugawa}
\affiliation{Quantum Transport Properties Group, National Institute for Materials Science, Tsukuba 305-0003, Japan}
\author{Shinya~Uji}
\affiliation{Quantum Transport Properties Group, National Institute for Materials Science, Tsukuba 305-0003, Japan}
\affiliation{Graduate School of Pure and Applied Sciences, University of Tsukuba, Tsukuba, Ibaraki 305-8577, Japan}
\author{Yoshiteru~Maeno}
\affiliation{Department of Physics, Graduate School of Science, 
Kyoto University, Kyoto 606-8502, Japan}

\email{chanchal@scphys.kyoto-u.ac.jp, maeno@scphys.kyoto-u.ac.jp}

\date{\today}

\begin{abstract}

Non-equilibrium steady state (NESS) conditions induced by DC current can alter the physical properties of strongly correlated electron systems (SCES). In this regard, it was recently shown that DC current can trigger novel electronic states, such as current-induced diamagnetism, which cannot be realized in equilibrium conditions. However, reversible \rev{control} of diamagnetism has not been achieved yet. Here, we demonstrate reversible \textit{in situ} \rev{control} between a Mott insulating state and a diamagnetic semimetal-like state by DC current in the Ti-substituted bilayer ruthenate \crto\ ($x=0.5 \%$). By performing simultaneous magnetic and resistive measurements, we map out the temperature vs current-density phase diagram in the NESS of this material. The present results open up the possibility of creating novel electronic states in a variety of SCES under DC current.

\end{abstract}

\maketitle

%%\section{Introduction}  
%\linenumbers

%\section{Introduction}
After the discovery of high $T_{\mathrm{c}}$ superconductivity \cite{bednorz_possible_1986}, an intensive amount of research has been performed on strongly correlated electron systems (SCES) in the vicinity of the Mott insulating state. A multitude of quantum phenomena have been found in these materials including metal--insulator transitions, ferromagnetism, and unconventional superconductivity \cite{tokura2014colossal, keimer2015quantum, nakatsuji_quasi-two-dimensional_2000}. In these studies, Mott insulating states have been controlled using various equilibrium parameters such as external pressure, chemical composition, epitaxial strain, electric field, and magnetic field \cite{alireza_evidence_2010, imada1998metal, yang2011oxide, nakamura_electric-field-induced_2013, kuwahara1995first}. More recently, various nonequilibrium stimuli have been actively investigated \cite{ju2014photoinduced, chong2017observation, guiot2013avalanche, aoki2014nonequilibrium, cao2018electrical}. Such nonequilibrium stimuli can drive SCES into exotic states that are not accessible by other means. A striking example is the insurgence of giant diamagnetism in the Mott insulator \cro\ upon the application of a DC current of a few mA \cite{sow2017current}. However, in \cro\ it is not possible to perform \textit{in situ} control of magnetism because of its high electrical resistivity. Thus it is essential to search for a new class of materials which allows us to switch \textit{in situ} electronic and magnetic states as well as to investigate detailed evolution of the material properties under current. Such materials provide unique opportunity to study SCES under nonequilibrium steady state (NESS) \rev{conditions} and to construct the nonequilibrium phase diagram towards clarifying the origin of the NESS-induced phenomena. 

An interesting candidate to achieve this goal is the bi-layered ruthenate \crto\/, for which a Mott metal-insulator transition can be sensitively controlled by tiny Ti substitution \cite{cao1997observation, kikugawa2010ca3ru2o7, ke2011emergent, tsuda2013mott}. Pure \CRO\/, in fact, has a metallic and antiferromagnetic ground state which changes to an insulating state upon only 0.5\% substitution of Ru$^{4+}$ with nonmagnetic Ti$^{4+}$ \cite{kikugawa2010ca3ru2o7}. This leads to an increase of the resistivity up to 8 orders of magnitude below \Ts\ = 50~K \cite{kikugawa2010ca3ru2o7, tsuda2013mott, ke2011emergent, zhu2016colossal}. Hard X-ray photo-emission spectroscopy revealed that Ti substitution induces the opening of a Mott-like gap below \Ts\/. However, unlike conventional Mott insulators, the resistivity below \Ts\ presents a weak temperature dependence \cite{tsuda2013mott}, suggesting a competition between disorder effects and Mott physics in the framework of Mott-Anderson transition \cite{shinaoka2010theory}. These considerations along with relatively low resistivity at low temperature make \crto\ an ideal system to explore the possibility of inducing NESS effects \textit{in situ}.

In this letter, we report the discovery of current-induced diamagnetism in \crto\ ($x=0.5 \%$) by performing simultaneous measurements of electrical resistivity and magnetization under DC electric current. Our results demonstrate \textit{in situ} \rev{control} of diamagnetism and open up new possibilities for controlling SCES under NESS conditions.

%%%%%%%%%%%%%%%%%%%%%%%%%%%%%%%%%%%%%%%%%%%%%%%%%%%%%%%%
\begin{figure}[b]
\begin{center}
\includegraphics[width=8.5cm]{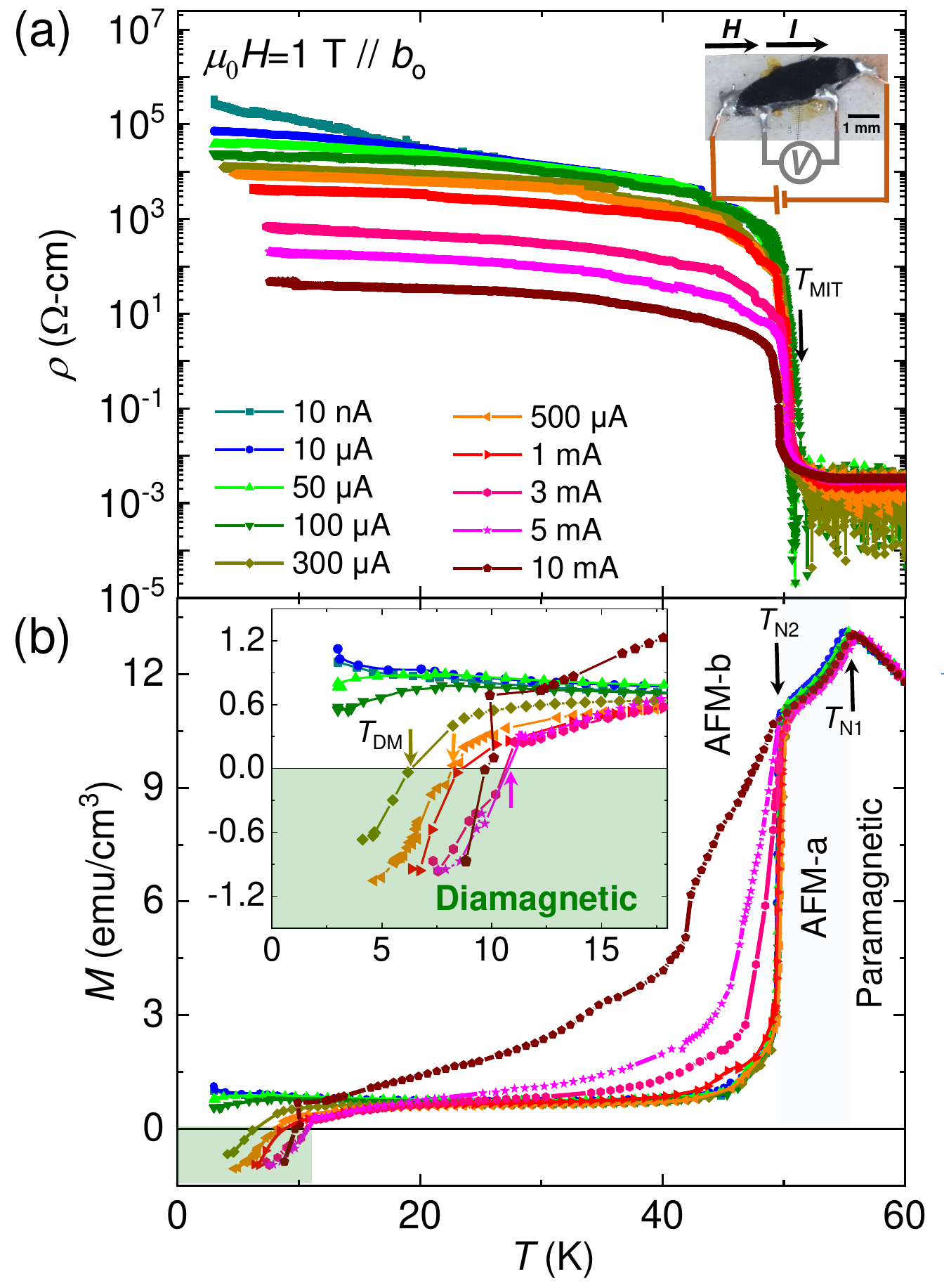}
\end{center}
\caption{
Current-induced diamagnetism in \crto\ ($x=0.5 \%$). Temperature dependence of (a) resistivity and (b) magnetization under various DC current. The photograph in (a) shows the measured single crystal \rev{(Sample \#1; dimensions of $3.3\times 0.8 \times 0.35$ mm$^3$)} along with a schematic description of the four-probe configuration. The inset in (b) is a zoom-in-view of the onset of diamagnetism. The arrows indicate the crossover temperature for diamagnetism \Td\/. 
\label{fig:main-data}
}
\end{figure}
%%%%%%%%%%%%%%%%%%%%%%%%%%%%%%%%%%%%%%%%%%%%%%%%%%%%%%%%

%\section{Experimental}

Single crystals of \crto\ (\textit{x} =0.5\%) were grown by a floating-zone method using an infrared image furnace (Canon Machinery, model SC-K15HD) in Kyoto University. We added extra RuO$_2$ both as self-flux and as compensation for RuO$_2$ evaporation \cite{mao2000crystal, perry2004systematic, zhou2005electronic, kikugawa2015single}. In this study, the starting materials CaCO$_3$ (99.999\%), RuO$_2$ (99.9\%), and TiO$_2$ (99.99\%) were mixed in the molar ratio of 3 : ($3-2x$) : $2x$. The feed rod (typical dimensions of 8 cm in length and 5 mm in diameter) was prepared by hydrostatic compression followed by sintering at 1000$^{\circ}$C in air for 2 hours. Single crystals were grown at a typical speed of 7 mm/h in a gas mixture of 90\% Ar and 10\% O$_2$ at a total pressure of 10 atm. Electron probe microanalysis (EPMA) with a commercial apparatus (JEOL, JXA 8500F) at NIMS revealed a homogeneous distribution of Ti in the grown crystals, with concentrations in good agreement with the nominal ones. The inset of Fig.~\ref{fig:main-data}(a) shows a photo of a typical  crystal \rev{(Sample \#1)} with four electrical contacts. \rev{Most of the data in this Letter was taken on this sample otherwise explicitly mensioned.}  A Laue pattern of this crystal is presented in the Supplementary Material \cite{SM}.
  
  We designed a sample holder compatible with a commercial SQUID magnetometer (Quantum Design, MPMS XL) for simultaneous measurements of transport and magnetic properties under DC current. The details of the design can be found in the Supplemental Material of Ref.~\cite{sow2017current}. We placed a bare-chip thermometer (Lake Shore, Cernox, CX-1050-BC-HT) close to the sample to monitor its temperature (Lake Shore, 335). The transport measurements were carried out with a current source (Keithley, 6221) and a nano-volt meter (Keithley, 2182) in a four-probe configuration. We used silver-epoxy (Epoxy Technology, EPO-TEK H20E) cured at 100$^{\circ}$C to provide electrical contacts with thin copper wires.
  
%%%%%%%%%%%%%%%%%%%%%%%%%%%%%%%%%%%%%%%%%%%%%%%%%%%%%%%%
\begin{figure}[b]
\begin{center}
\includegraphics[width=8.5cm]{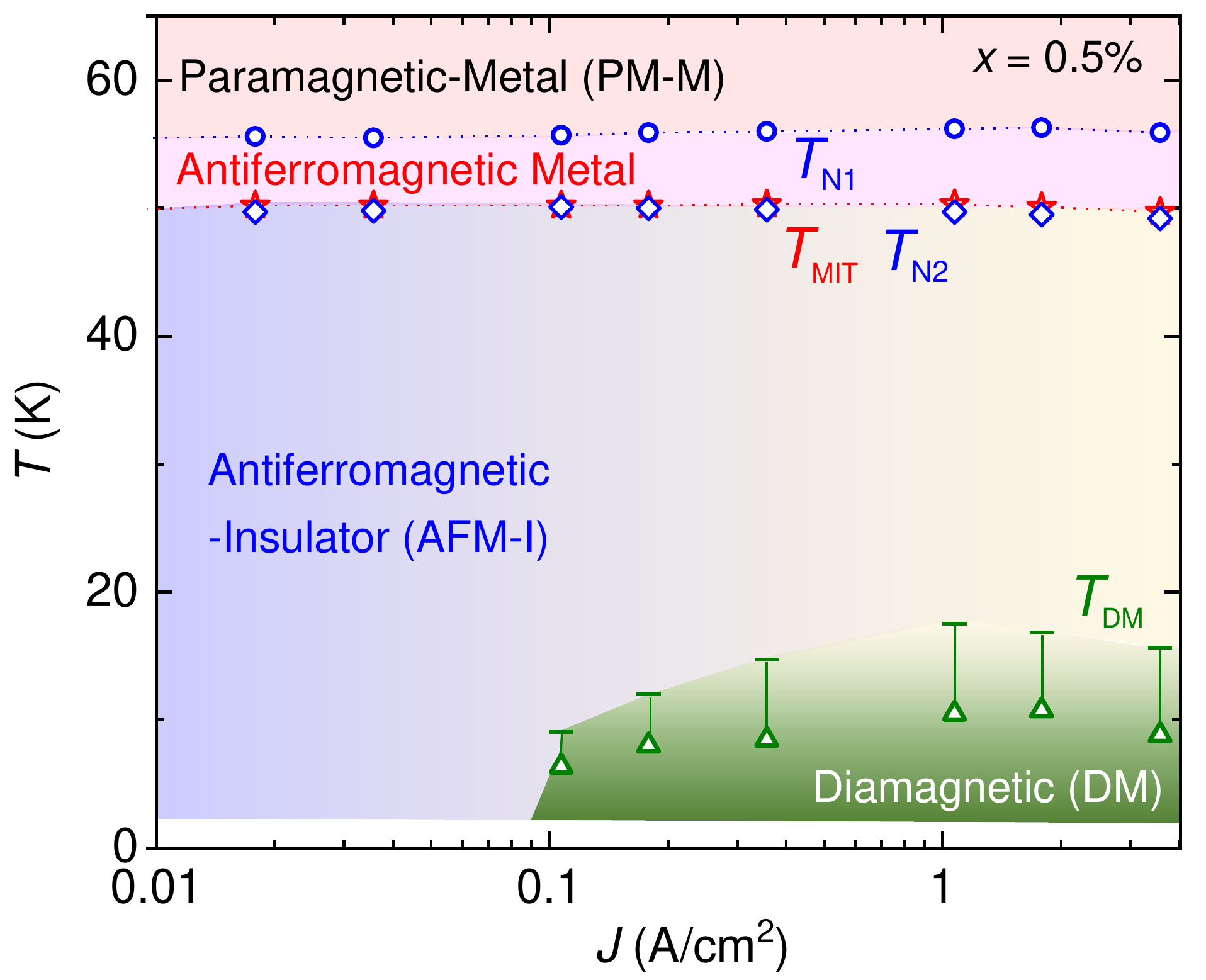}
\end{center}
\caption{
Temperature vs current-density phase diagram of \crto\ ($x=0.5 \%$). \Ta\/, \Tb\, and \Ts\ all show negligible changes with current. The antiferromagnetic insulating phase (blue) evolves into a more conducting state (yellow) with increasing current density. At temperatures below 10~K a diamagnetic phase (green) appears for $J>0.1 \,\mathrm{A/cm}^2$ with persisting AFM-$b$ ordering. \rev{The triangles represent \Td\ taken from Fig.~\ref{fig:main-data}, while the error bars are evaluated by estimating the real sample temperature due to Joule heating  from another test experiment (see Fig.~\ref{fig:thermometer-on-top}).}
\label{fig:phase-diagram}
}
\end{figure}
%%%%%%%%%%%%%%%%%%%%%%%%%%%%%%%%%%%%%%%%%%%%%%%%%%%%%%%%

%\section{Results and Discussion}    
%\subsection{Diamagnetism under current}

Figure~\ref{fig:main-data} presents results of simultaneous measurements  of resistivity and magnetisation under various DC currents. We used an external field $\mu_{\mathrm 0}H=1$~T applied along the orthorhombic $b_{\mathrm o}$ crystallographic axis and parallel to the current direction. As shown in Fig.~\ref{fig:main-data}(a), the resistivity above $50\,\mathrm{K}$ is metallic with a value of the order of $10^{-3}\,\Omega\,\mathrm{cm}$ in agreement with the previous report \cite{tsuda2013mott}. Below \Ts\ $\sim50\,\mathrm{K}$, the resistivity shows a sharp step-like increase, reaching a value of $10^{5}\,\Omega\,\mathrm{cm}$ at $2\,\mathrm{K}$ and $10~\mu$A. This value is 8 orders of magnitude higher than that in the high-temperature metallic phase. With increasing DC current, the resistivity in the metallic phase shows negligible variation (see also Fig.~\ref{fig:standard-data}a). However, a large change occurs in the insulating phase, where the resistivity is gradually reduced with increasing current, becoming almost 4 orders of magnitude smaller at $10\,\mathrm{K}$ and 10~mA. \rev{Correspondingly, the current-voltage characteristic in this insulating state is highly non-linear, as shown in Fig.~\ref{fig:I-V}. We note that the resistivity curves show a shift of up to 2 K around \Ts\/, which is probably extrinsic considering possible Joule heating and slight variations in the sample cooling (further details in Fig.~\ref{fig:thermometer-on-top}).} The magnetization curves in Fig.~\ref{fig:main-data}(b) lie on top of each other in the high temperature region and show a peak at the Neel temperature \Ta\,=56~K, corresponding to the onset of the antiferromagnetic AFM-$a$ phase. The magnetization exhibits a sharp decrease at \Tb\,=50~K corresponding to the onset of the AFM-$b$ phase \cite{peng2013quasi}. We note that the AFM-$b$ order and the metal-to-insulator transition occur at the same temperature (\Tb\ = \Ts\ = 50~K) and are accompanied by a structural transition with reduction in the \textit{c} axis \cite{yoshida2005crystal}. The increase in the magnetization with DC current below \Tb\ is attributable to a decrease in the stiffness \rev{\cite{cullity2011introduction}} of the magnetic structure.

%\subsection{diamagnetism found}
The most interesting change is observed \rev{upon further cooling}: $M$ becomes negative below the crossover temperature \Td\ for applied current above $300~\mu$A indicating the appearance of a novel diamagnetic state. As the current is increased, the crossover temperature increases as shown in the inset of Fig.~\ref{fig:main-data}(b).

%\subsection{Phase diagram}
We use the measured transition temperatures to construct the $T$--$J$ phase diagram for the various magnetic phases as a function of current density in Fig.~\ref{fig:phase-diagram}. The diamagnetic phase appears in the region above $J=0.1\,\mathrm{A/cm}^2$ and below  \rev{10~K, where we used the geometric conversion relation $I$ (mA) $=2.8\,J(\mathrm{A/cm}^2)$ as deduced from the sample cross section.} Importantly, the emerging diamagnetic state in \crto\ coexists with the AFM order. This is evidenced by the presence of the sharp transition at \Tb\ in Fig.~\ref{fig:main-data}(b) for all the explored values of DC current. It is interesting to compare the current-induced diamagnetism of \crto\ with what is observed in \cro\/. In the case of \cro\/, the AFM order is completely suppressed with current when the diamagnetism emerges \cite{sow2017current, bertinshaw2018unique}. Thus, the coexistence of diamagnetism with AFM order is a characteristic feature of \crto\/. As another important difference from \cro\/, the lower resistivity of \crto\ allows us to access the full $T$--$J$ phase diagram and perform detailed investigations of current-induced phenomena. \rev{By measuring the magnetization as a function of current density at 6 K (Fig.~\ref{fig:curr-dep}), we find out that the emergence of diamagnetism is a crossover rather than a phase transition. The resistivity also shows a continuous decrease as a function of current, but its slope changes at about 0.1 $\mathrm{A/cm}^2$, a possible signature of a structural and/or electronic change which is beyond the scope of the present work. The temperature dependence of the resistivity is rather continuous at \Td\/, indicating again the absence of a phase transition in the diamagnetic state.}

%%%%%%%%%%%%%%%%%%%%%%%%%%%%%%%%%%%%%%%%%%%%%%%%%%%%%%%%
\begin{figure}
\begin{center}
\includegraphics[width=8.5cm]{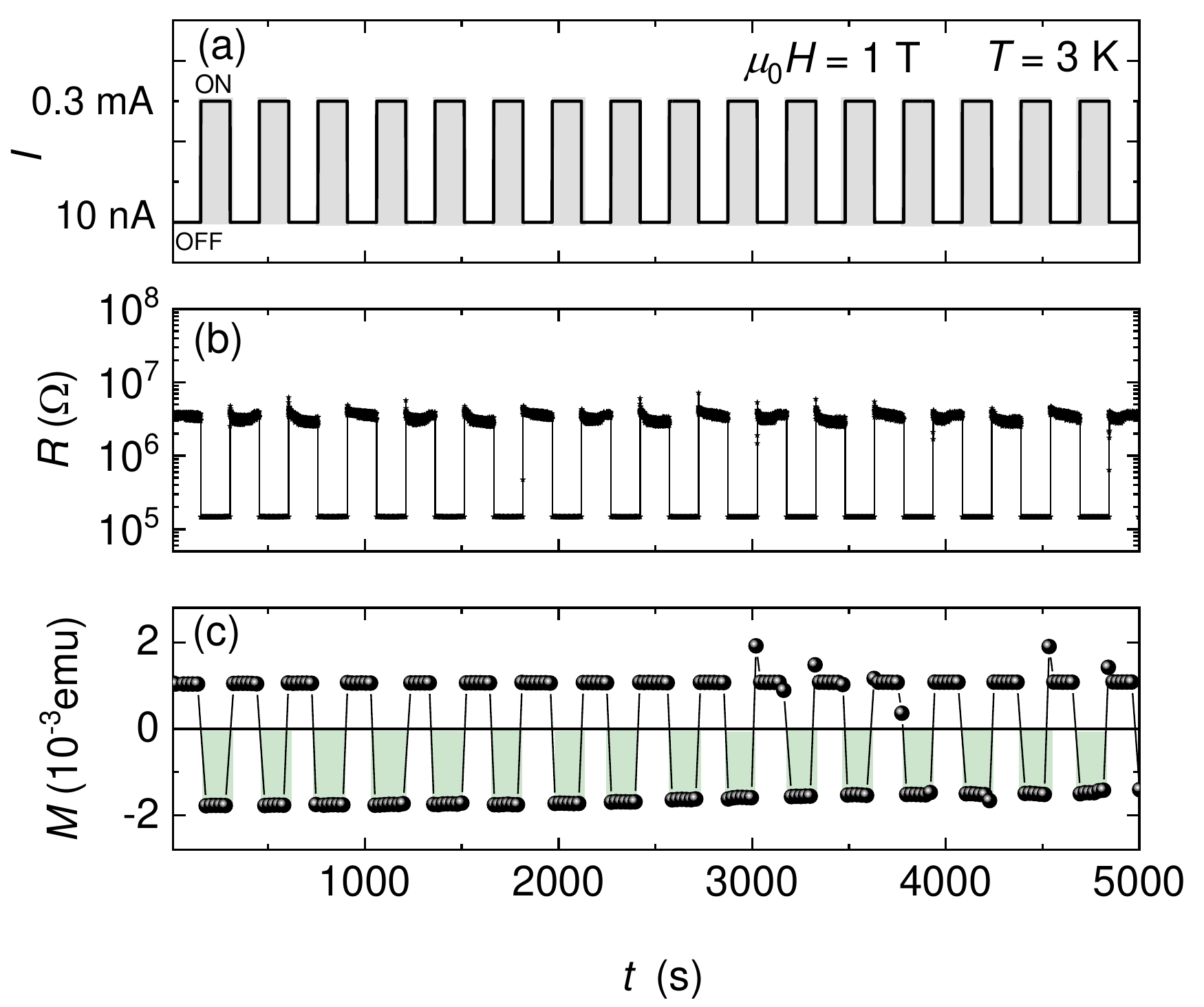}
\end{center}
\caption{
\rev{\textit{In-situ} control of diamagnetism with current in \crto\ ($x=0.5 \%$). (a) Cycling the current between the OFF and ON state induces (b) a change between the high- and low- resistance state and (c) between the positive and negative magnetization (diamagnetic state). Due to the different Joule heating in the two states, sample temperature varied between 2 and 4 K in each cycle (see Fig.~\ref{fig:switching-details}).} 
\label{fig:switching-demo}
}
\end{figure}
%%%%%%%%%%%%%%%%%%%%%%%%%%%%%%%%%%%%%%%%%%%%%%%%%%%%%%%%

%\subsection{In situ Switching}
We now demonstrate that reversible \rev{control} of resistance and magnetization can be achieved by changing the applied DC current. As shown in Fig.~\ref{fig:switching-demo}(a), we alternate the DC current between an OFF state (10~nA) and an ON state (300~$\mu$A) while simultaneously measuring the resistance (Fig.~\ref{fig:switching-demo}(b)) and the magnetization (Fig.~\ref{fig:switching-demo}(c)) with a constant applied magnetic field $\mu_{\mathrm 0}H=1$~T. The magnetization \rev{changes} between positive and negative as the current is cycled between the OFF and ON states. This is, to the best of our knowledge, the first demonstration of reversible \textit{in situ} \rev{control} of diamagnetism by DC current. The resistance concomitantly changes from a \rev{high- to a low-resistance state. We demonstrated that such alternation between the states is reproducible over more than $10^{5}$ repetition cycles below 10 K (Fig.~\ref{fig:switching-hystogram} (a-c)). This suggests that the occurrence of diamagnetism has mainly electronic origins, and possible slower and irreversible processes, such as chemical reaction or oxygen diffusion, play a negligible role.} By repeating similar experiments with a reversed magnetic field of $-1$~T (Fig.~\ref{fig:switching-details}(a)) or with reversed current of $-0.3$ mA (Fig.~\ref{fig:switching-details}(b)), we confirmed \rev{that the diamagnetism occurs only in the current ON state} irrespective of the current direction. This fact clarifies that the magnetic field created by the applied current via Ampere's law is not the origin of the observed negative magnetization phenomenon. \rev{We also note that the direction of current has no effect on the value of the resistance (Fig.~\ref{fig:switching-details}), as also shown in the symmetric $I-V$ characteristics (Fig.~\ref{fig:I-V})}. Concerning the time scale of the switching, we found that the resistive \rev{change} occurs faster than 0.1 s and magnetic \rev{change} faster than 20 s, which is limited by the slow repetition rate of our SQUID measurements. Further experiments with much shorter data acquisition rate are needed to reveal the dynamics of this \rev{change} on faster time scales.  

%%%%%%%%%%%%%%%%%%%%%%%%%%%%%%%%%%%%%%%%%%%%%%%%%%%%%%%%
\begin{figure}
\begin{center}
\includegraphics[width=8.5cm]{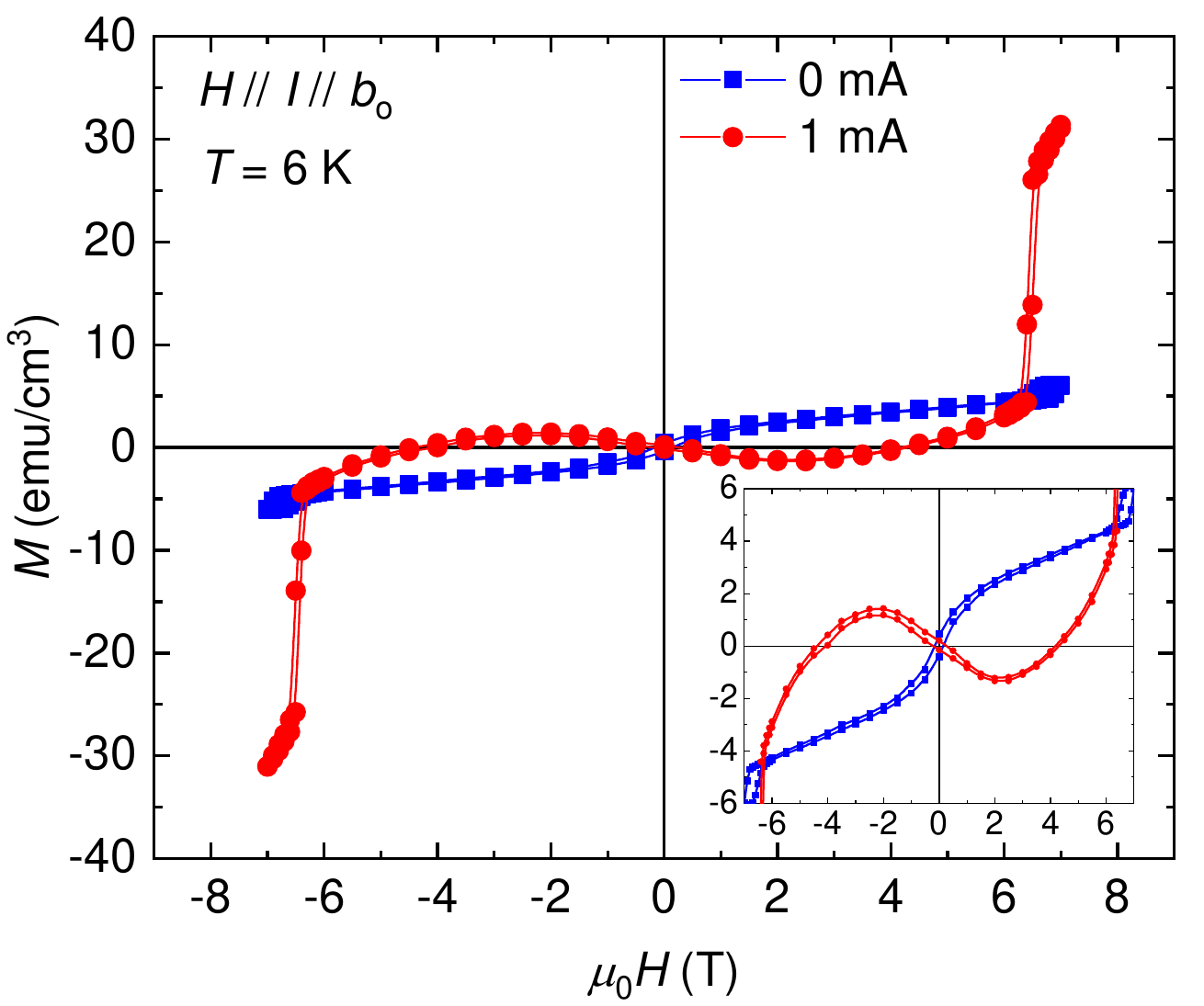}
\end{center}
\caption{
Magnetization vs magnetic field (\textit{M}--\textit{H}) curves for \crto\ ($x=0.5 \%$) with and without DC current. Both field-up and field-down data are shown. The sudden increase of magnetization under current indicates the occurrence of a metamagnetic transition. The inset shows an enlarged view near $M=0$. 
\label{fig:h-dep}
}
\end{figure}
%%%%%%%%%%%%%%%%%%%%%%%%%%%%%%%%%%%%%%%%%%%%%%%%%%%%%%%%
%\subsection{H dependence}
We now compare the field dependence of the magnetization with and without current. For this measurement both $H$ and $I$ are applied parallel to the $b_{\mathrm o}$ axis. As shown in Fig.~\ref{fig:h-dep}, for $I=0$ the magnetization exhibits a positive slope at all fields. The presence of a weak hysteresis and a change in the slope at $\pm$1 T are consistent with the presence of canted moments in the AFM-$b$ phase \cite{peng2013quasi}. For $I=1\,\mathrm{mA}$, $M(H)$ has a negative slope up to 2~T, indicating  negative magnetic susceptibility, characteristic of a diamagnetic state. We comment that this is essentially different from negative magnetization arising from other origins, such as current-induced flipping of magnetization reported in the ferrimagnet YVO$_3$ \cite{ren_temperature-induced_1998} or magneto-electric effects in multiferroics \cite{eerenstein2006multiferroic}. With further increase of the magnetic field, the magnetization slope changes to positive for $\mu_{\mathrm 0}H>2.5$~T, after which also $M$ becomes positive. At 6.5~T the magnetization suddenly jumps by a factor of 7, indicating the occurrence of a metamagnetic transition, where the AFM-$b$ ordered phase changes into a field-induced ferromagnetic state \cite{ohmichi2004colossal, peng2016magnetic}. As an effect of this transition, the resistance exhibits a flattening around 6.5~T (Fig.~\ref{fig:MR-data}(a)). The existence of the metamagnetic transition provides an additional piece of evidence that the AFM-$b$ order persists under current. We note that the metamagnetic transition without current was not observed in our sample up to 7~T. Thus the electric current promotes the metamagnetic transition, consistent with the reduction in the magnetic stiffness discussed above.

Because of the coexistence of diamagnetism and AFM-$b$, the magnetization signal contains both contributions. We attempt to isolate the diamagnetic contribution, which appears only under current, by evaluating $\Delta M=M(1\,\mathrm{mA})-M(0\,\mathrm{mA})$ as a function of $\mu_{\mathrm 0}H$ (Fig.~\ref{fig:MR-data}(b)). The actual diamagnetic contribution would be greater because the AFM contribution depends somewhat on $I$ due to weakening of the magnetic stiffness even at 6~K. We find $\Delta M \sim -2.3$~emu/$\mathrm{cm}^3$ up to 1~T, a value comparable to that in \cro\ under current and larger than those in the strong diamagnets bismuth or graphite \cite{sow2017current}. 

%\subsection{Mechanism}
In the rest of this Letter, we discuss the possible origin of the current-induced diamagnetism in \crto\/. In its sister compound \cro\/, similar current-induced diamagnetism has been observed and attributed to the Landau diamagnetism of light-mass quasiparticles generated by the partial Mott gap closing \cite{sow2017current}. \cro\ is a Mott insulator where $d_{{xy}}$ is fully occupied due to the flattening of the RuO$_6$ octahedra, and the $d_{{yz}}$ and $d_{{zx}}$ are half filled and exhibit the opening of a Mott-gap. Under DC current, it is experimentally known that the Mott gap is reduced, possibly originating from the reduction of the effective electron correlation $U$ \cite{okazaki_current-induced_2013}. When $U$ is reduced, it is proposed that the Mott gap first closes where the electronic state has a strong 2D character due to $d_{{yz}}$/$d_{{zx}}$ hybridization. This strongly $k$-dependent gap closing results in highly dispersive (light mass) quasiparticles, which leads to strong diamagnetism.

The ground state of pure \CRO\ has a metallic character with a possible small gap in some parts of the Fermi surface \cite{kikugawa2010ca3ru2o7, yoshida2004quasi}. ARPES results indicate that the Fermi surface originates from quasi-1D $d_{{yz}}$ and $d_{{xz}}$ bands \cite{baumberger2006nested}. By Ti substitution, \crto\ becomes Mott insulating with gap opening in the whole Fermi surface \cite{tsuda2013mott}. Upon application of a DC current we expect the effective correlation to be weakened and hence the Mott gap to be reduced similarly to \cro\ \cite{okazaki_current-induced_2013}. This picture is corroborated by the observed reduction of the resistivity shown in Fig.~\ref{fig:main-data}(a). In this picture, the strong diamagnetism in \crto\ is ascribable to thermally excited quasiparticles with very light effective mass emerging when the Mott gap is about to close.

%\subsection{Diamagnetic contribution}
Important novel aspects are found in the diamagnetic state of \crto\/. Firstly, the current-induced diamagnetism occurs in a insulating state with much lower resistivity compared to \cro\/. Secondly, diamagnetism in \crto\ coexists with the AFM ordering. This important difference may originate from the difference in the nearest-neighbor spin alignment: it is ferromagnetic within a Ru-O bilayer in \crto\ whereas it is G-type AFM in \cro\ \cite{yoshida2005crystal}. These differences should be taken into account when constructing a realistic model to describe the diamagnetism in \crto\/. In particular, possible interaction between the itinerant diamagnetism and localized AFM is an interesting issue to be explored. It is intriguing that current-induced diamagnetism is discovered in systems with different transport and magnetic characteristics, implying that it may arise in other SCES under the NESS condition.

%\section{Conclusions, Future perspectives}
To summarize, we demonstrated that the insulating state of \crto\ system can be electronically controlled by flowing electric current. Most importantly, the antiferromagnetic insulating state can be reversibly \rev{changed} \textit{in situ} to a diamagnetic state with a lower resistivity. The basic origin of the diamagnetism in \crto\ is believed to be caused by the light-mass carriers generated in the NESS. The present work reveals that DC current is a new control tool to \rev{control} the physical properties of a variety of SCES, with the possibility of inducing novel states which are not accessible by conventional control parameters.

%\nolinenumbers

%\section*{Acknowledgements}
 This work was supported by JSPS Grant-in-Aids KAKENHI Nos. JP26247060, JP15H05852, JP15K21717, and JP17H06136, as well as JSPS Core-to-Core program). N. K. and S. U. acknowledge the support from JST-Mirai Program (No. JPMJMI18A3) in Japan. We acknowledge fruitful discussions with Takashi Oka, Sota Kitamura, Kazuhiko Kuroki, Teppei Yoshida and Masatoshi Imada.

\bibliography{CRO}
\clearpage

\clearpage
\appendix
\renewcommand{\theequation}{S\arabic{equation}}
\setcounter{equation}{0}
\renewcommand{\thefigure}{S\arabic{figure}}
\setcounter{figure}{0}
\renewcommand{\thetable}{S\arabic{table}}
\setcounter{table}{0}
\renewcommand{\thepage}{S\arabic{page}}
\setcounter{page}{1}
\renewcommand{\thesubsection}{S\arabic{subsection}}
\setcounter{equation}{0}
\begin{center}

\vspace{0.3cm}

{\large Supplementary Materials for\\[0.3cm]
\bfseries{\rev{\textit{In situ} control of diamagnetism by electric current in \crto }}}

% Place the author information here.  Please hand-code the contact
% information and notecalls; do *not* use \footnote commands.  Let the
% author contact information appear immediately below the author names
% as shown.  We would also prefer that you don't change the type-size
% settings shown here.

\vspace{0.5cm}

{\large
Chanchal~Sow$^{1\ast}$, Ryo~Numasaki$^{1}$, Giordano~Mattoni$^{1}$, Shingo~Yonezawa$^{1}$, Naoki~Kikugawa$^{2}$, \\
Shinya~Uji$^{2,3}$, \& Yoshiteru~Maeno$^{1\dagger}$\\

\vspace{0.4cm}

\noindent
\small
 {$^{1}$Department of Physics, Graduate School of Science,}\\  
\small
 {Kyoto University, Kyoto 606-8502, Japan }\\[0.2cm] 
\small
 { $^2$Quantum transport properties group, \\ 
 National Institute for Materials Science,} \\

\small
 {  Tsukuba 305-0047, Japan }\\[0.2cm] 
\small
 { $^3$Graduate School of Pure and Applied Sciences,}\\

\small
 {University of Tsukuba, Tsukuba, Ibaraki 305-8577, Japan}\\[0.2cm] 
 
\vspace{0.4em}
\noindent
\normalsize $^{\ast}$e-mail: chanchal@scphys.kyoto-u.ac.jp, $^{\dagger}$e-mail: maeno@scphys.kyoto-u.ac.jp

}

\date{\normalsize \it  \today}

\end{center}

\vspace{0.5cm}

\section*{Supplementary Figures}
%%%%%%%%%%%%%%%%%%%%%%%%%%%%%%%%%%%%%%%%%%%%%%%%%%%%%%%%
\begin{figure}[b]
\begin{center}
\includegraphics[width=8cm]{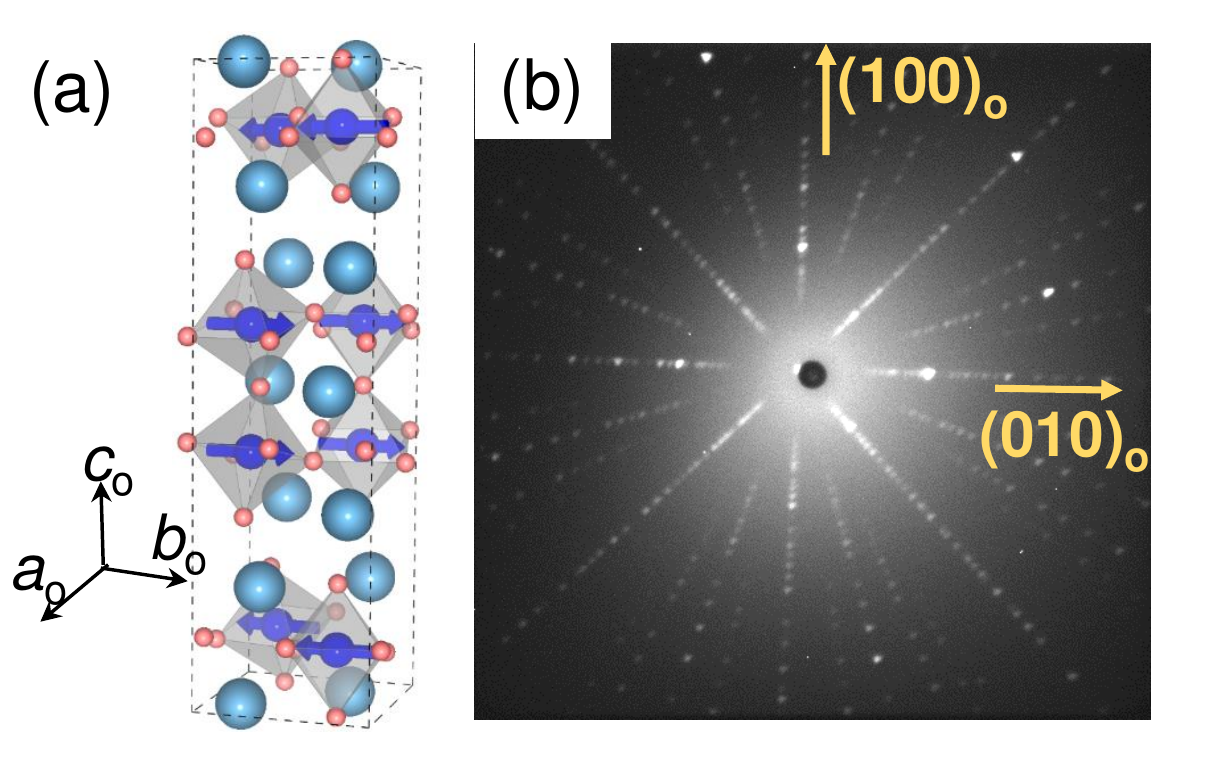}
\end{center}
\caption{
(a) Schematic crystal structure of \CRO\  \cite{peng2013quasi}. The orthorhombic crystalline axes and the magnetic structure in the antiferromagnetic AFM-\textit{b} phase are indicated. (b) Laue photograph of the \crto\ crystal used for the transport and magnetic measurements of Fig.~\ref{fig:main-data} \rev{(Sample \#1)}.  
\label{fig:crystal-structure}
}
\end{figure}
%%%%%%%%%%%%%%%%%%%%%%%%%%%%%%%%%%%%%%%%%%%%%%%%%%%%%%%%
\begin{figure}[b]
\begin{center}
\includegraphics[width=8cm]{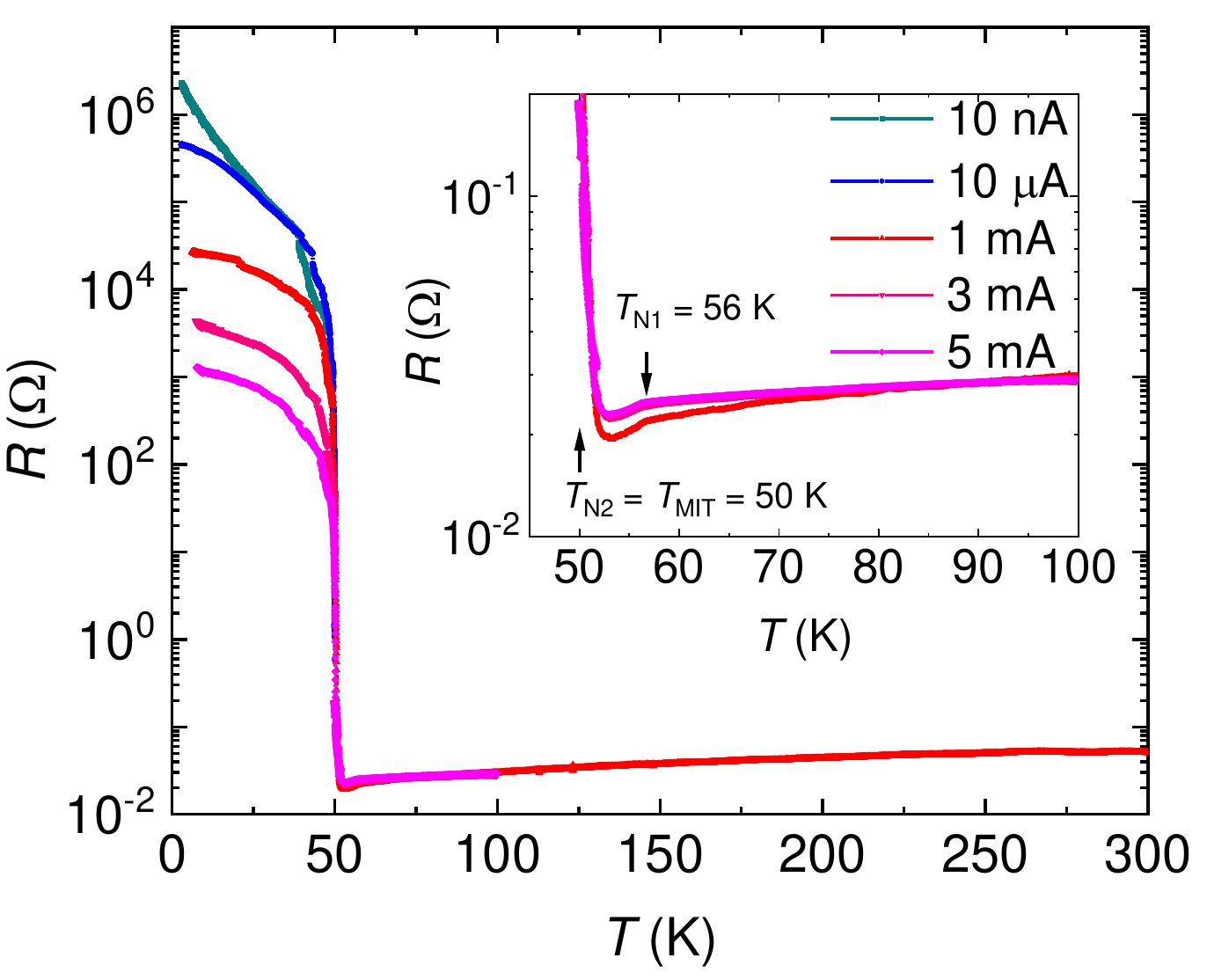}
\end{center}
\caption{
 Temperature dependence of the resistance of \crto\ ($x=0.5\%$)  \rev{(Sample \#1)} measured under various values of DC current at 1~T. The inset shows a zoom-in-view. These measurements were carried out with a constant positive current, the small current-dependent shift of the curves in the metallic phase is most likely due to uncompensated voltage offsets. 
\label{fig:standard-data}
}
\end{figure}

%%%%%%%%%%%%%%%%%%%%%%%%%%%%%%%%%%%%%%%%%%%%%%%%%%%%%%%%
\begin{figure*}
\begin{center}
\includegraphics[width=14cm]{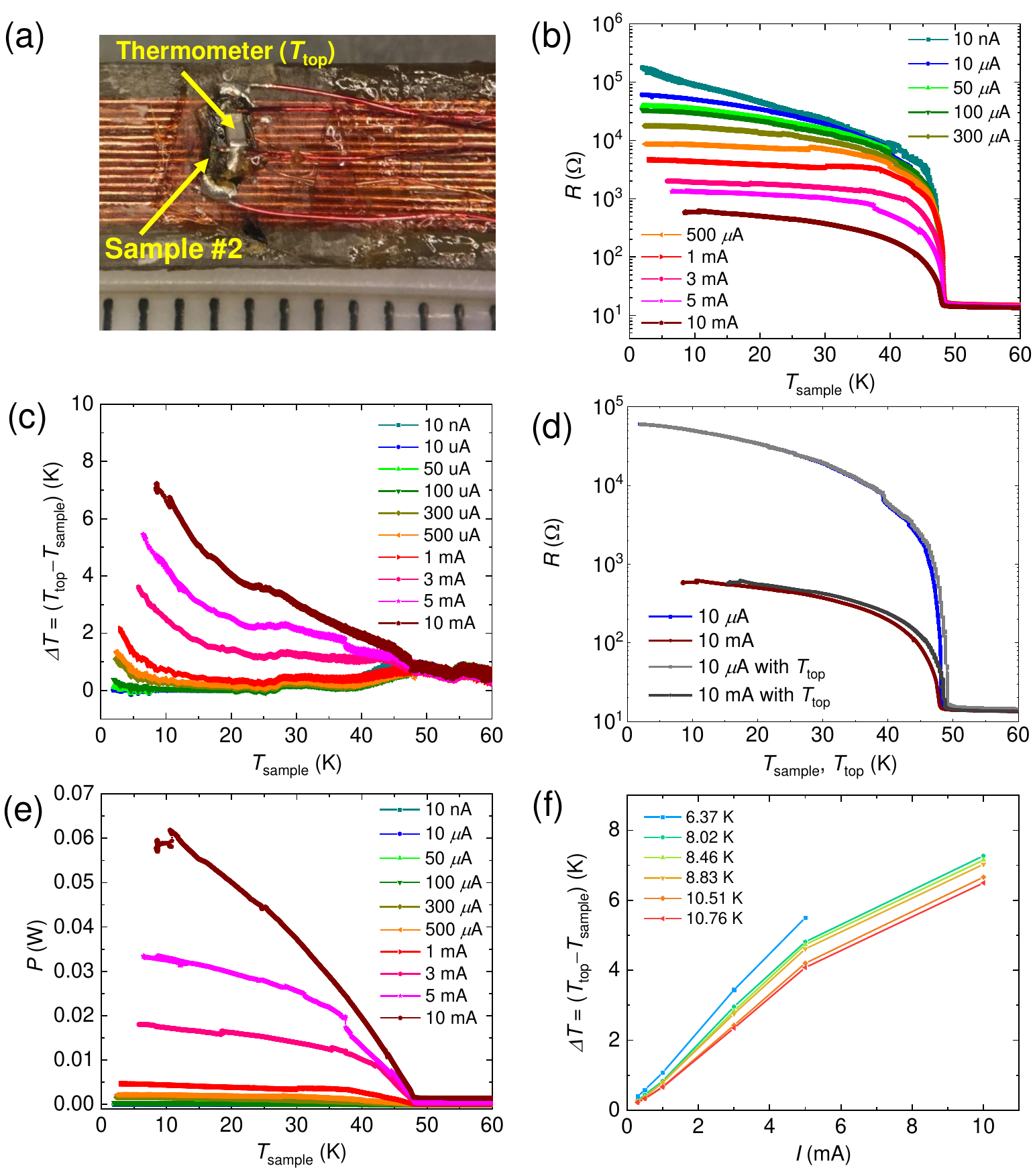}
\end{center}
\caption{
\rev{(a) Photograph of the resistive RuO$_x$ thick-film thermometer (KOA, RK73K1E) fixed with a low-temperature thermally-conducting glue (GE varnish) on top of a \crto\ ($x=0.5 \%$) crystal (Sample \#2 of size $3\times 0.9 \times 0.5$ mm$^3$) equivalent to the one used in the main text. The scale in the bottom is 1 mm per division. (b) Additional two-probe resistance as a function of temperature measured at various currents on Sample \#2 which are consistent with the trend reported in Fig.~\ref{fig:main-data}(a) of the main text. Here the temperature is measured by the same thermometer used in the main text ($T_{\mathrm {sample}}$) which is 2-cm away from the sample. (c) temperature difference between readings of the additional thermometer ($T_{\mathrm {top}}$) and $T_{\mathrm {sample}}$. (d) Comparison of sample resistance plotted against $T_{\mathrm {top}}$ and $T_{\mathrm {sample}}$. (e) Total power dissipated by the sample measured at various currents. (f) Temperature difference $\Delta T=(T_{\mathrm {top}}-T_{\mathrm {sample}})$ measured as a function of current at various temperatures corresponding to \Td\ in Fig.~\ref{fig:phase-diagram}. This $\Delta T$ gives the error bars in \Td\ of Fig.~\ref{fig:phase-diagram}. 
} 
\label{fig:thermometer-on-top}
}
\end{figure*}
%%%%%%%%%%%%%%%%%%%%%%%%%%%%%%%%%%%%%%%%%%%%%%%%%%%%%%%%
\begin{figure}
\begin{center}
\includegraphics[width=8cm]{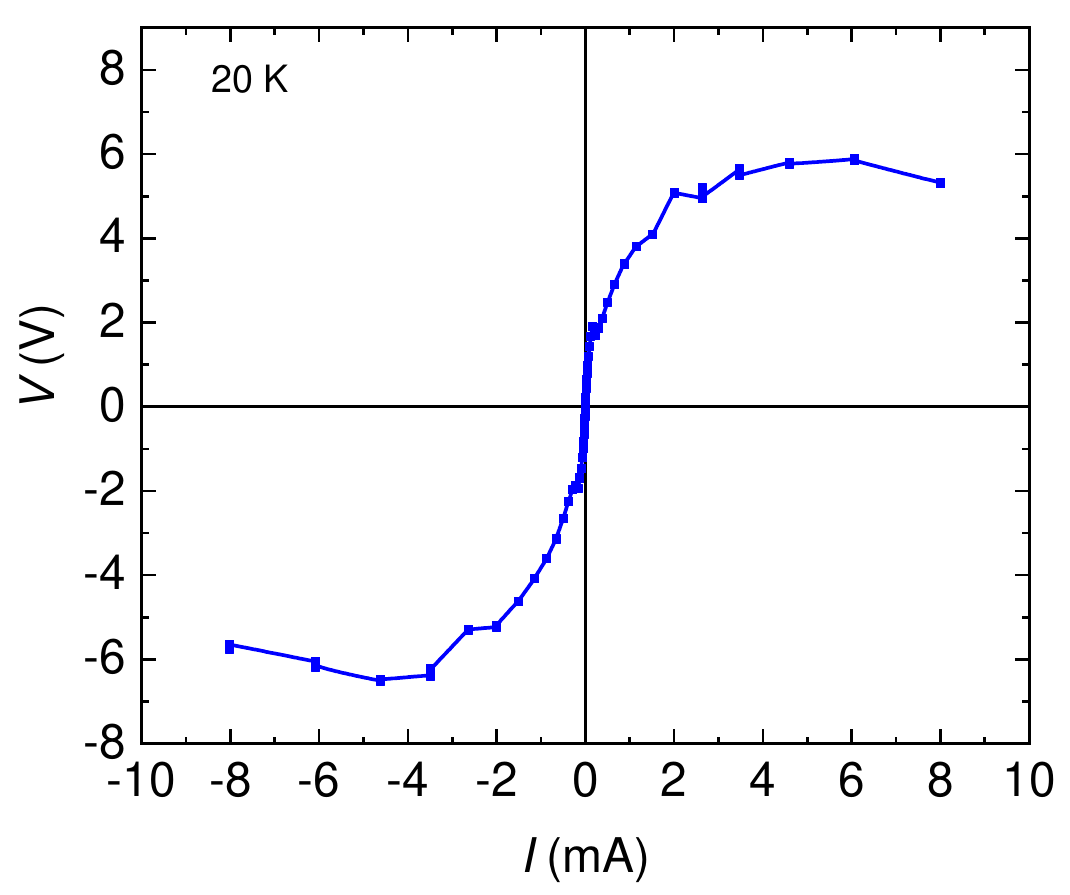}
\end{center}
\caption{
\rev{$I$-$V$ characteristics measured at 20 K for Sample \#2.} 
\label{fig:I-V}
}
\end{figure}

%%%%%%%%%%%%%%%%%%%%%%%%%%%%%%%%%%%%%%%%%%%%%%%%%%%%%%%%

%%%%%%%%%%%%%%%%%%%%%%%%%%%%%%%%%%%%%%%%%%%%%%%%%%%%%%%%
\begin{figure}
\begin{center}
\includegraphics[width=8cm]{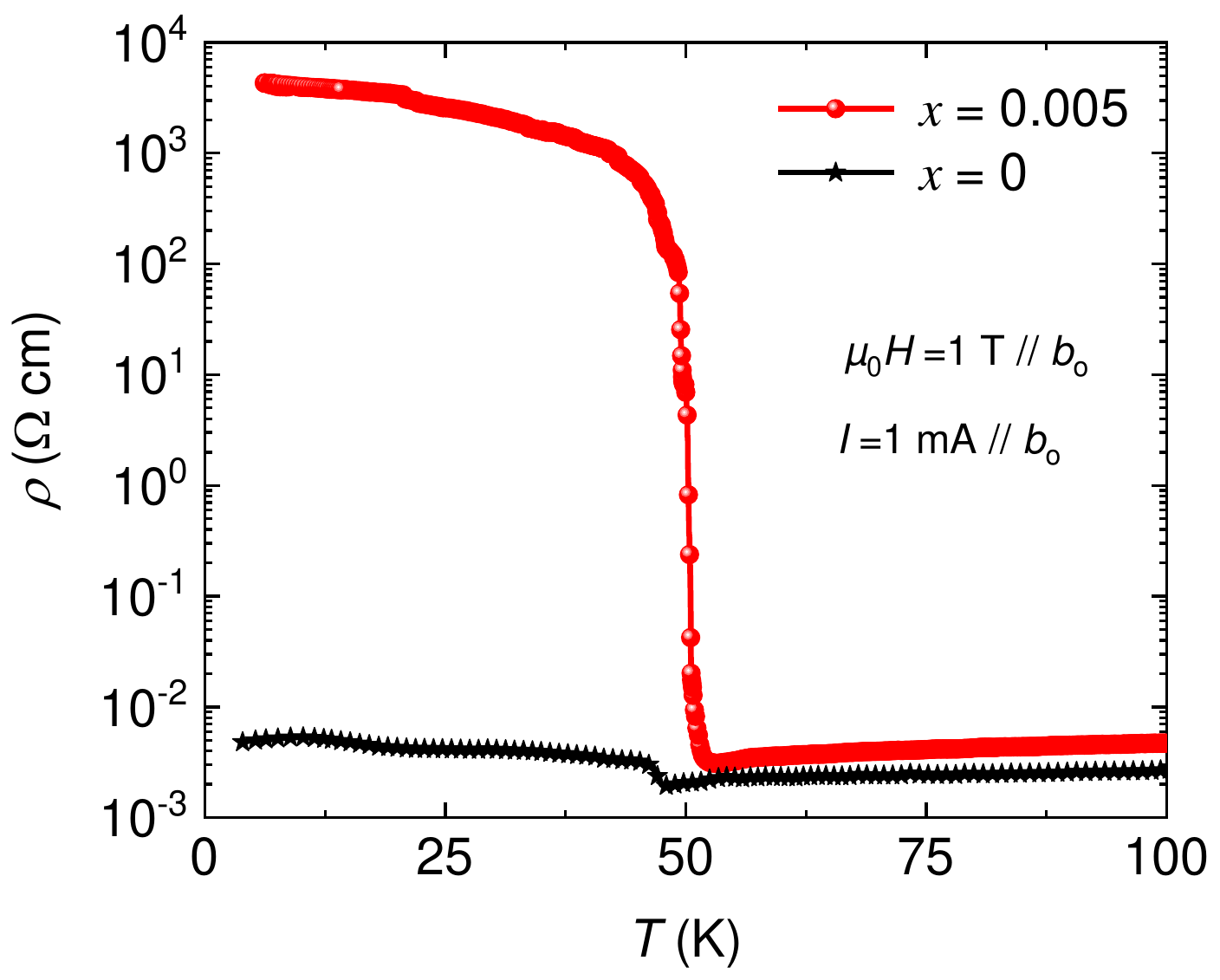}
\end{center}
\caption{
Resistivity vs temperature  measured at 1 mA for pure \CRO\ in comparison with that for a \crto\ ($x=0.5\%$) \rev{sample \#1}. 
\label{fig:resistance-comparison}
}
\end{figure}

%%%%%%%%%%%%%%%%%%%%%%%%%%%%%%%%%%%%%%%%%%%%%%%%%%%%%%%%
\begin{figure}
\begin{center}
\includegraphics[width=8cm]{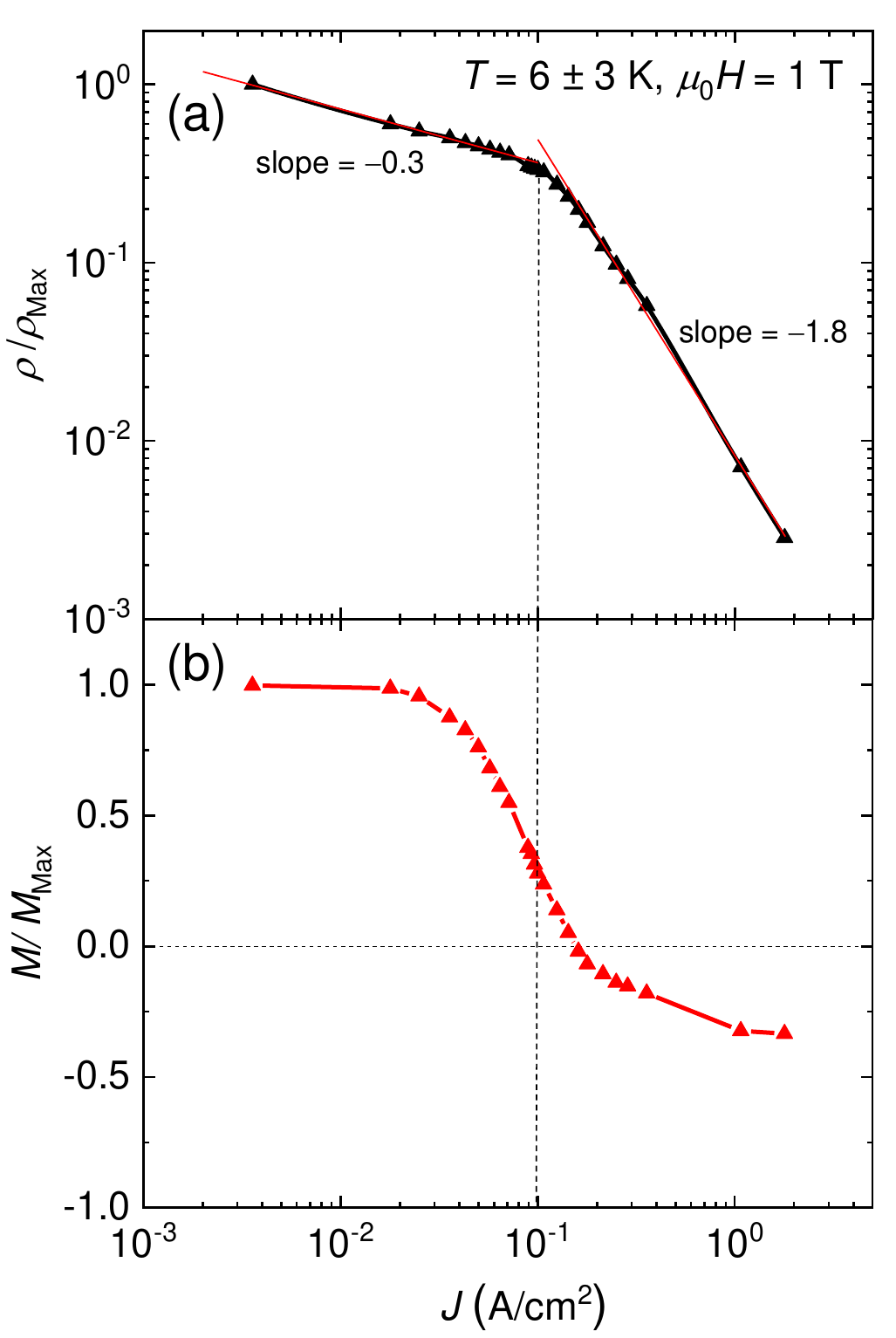}
\end{center}
\caption{
Current-density dependence of (a) resistivity and  (b) magnetization in \crto\ ($x=0.5\%$) \rev{sample \#1}. The temperature increases from 3 K to 9 K due to Joule heating under high current.   
\label{fig:curr-dep}
}
\end{figure}

%%%%%%%%%%%%%%%%%%%%%%%%%%%%%%%%%%%%%%%%%%%%%%%%%%%%%%%%
\begin{figure*}
\begin{center}
\includegraphics[width=8cm]{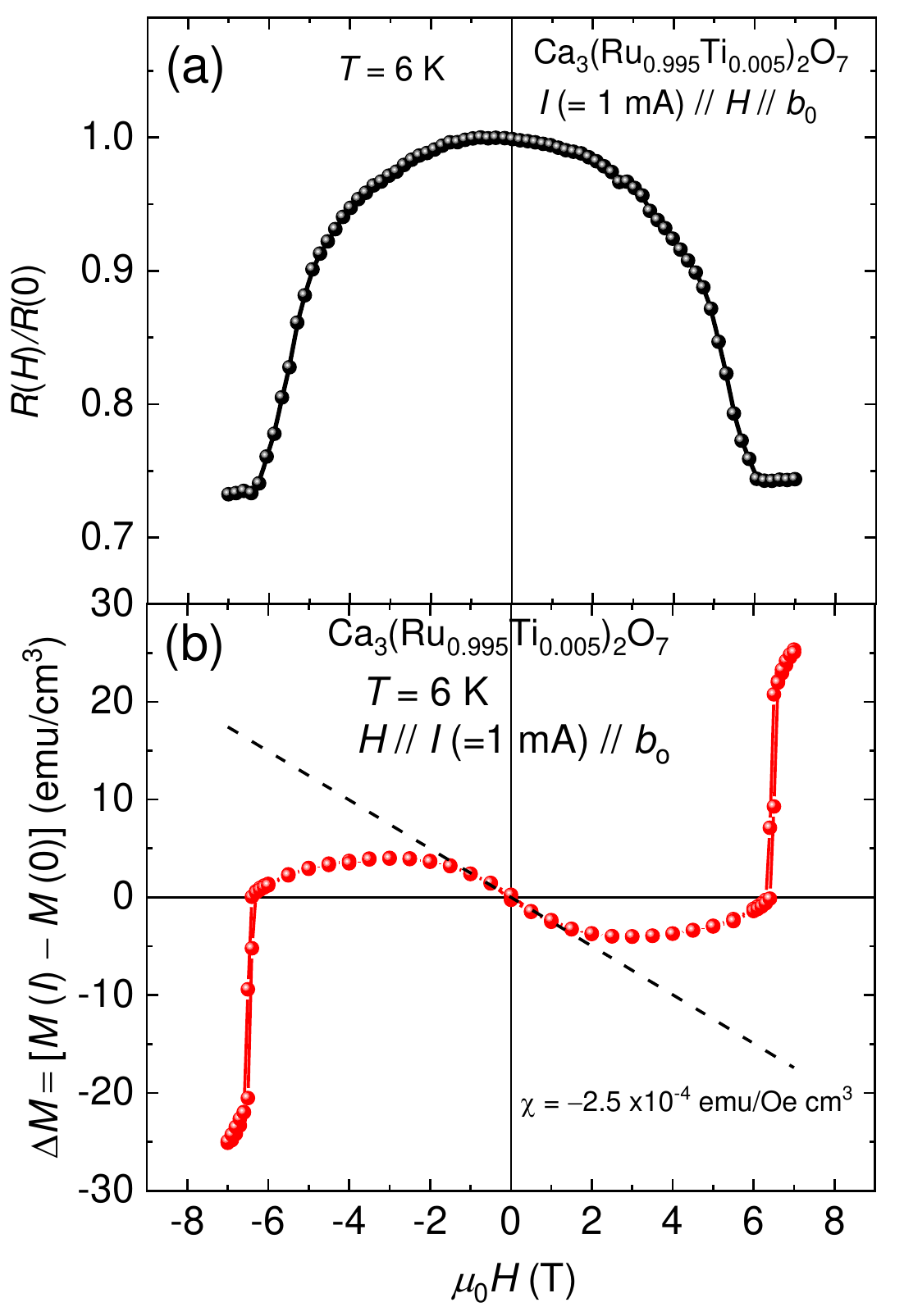}
\end{center}
\caption{
(a) Normalized resistance vs magnetic field at 6 K for \crto\ with $x=0.5\%$ \rev{sample \#1}. The flattening above 6.5~T corresponds to the metamagnetic transition. (b) Change in the magnetization ($\Delta M$) due to the current flow as a function of applied magnetic field. The metamagnetic transition occurs at 6.5 T and the black line is a linear fit within $\pm1$~T yielding a diamagnetic susceptibility of $\chi= -2.5 \times$ 10$^4$ emu/Oe cm$^3$.   
\label{fig:MR-data}
}
\end{figure*}
%%%%%%%%%%%%%%%%%%%%%%%%%%%%%%%%%%%%%%%%%%%%%%%%%%%%
\begin{figure*}
\begin{center}
\includegraphics[width=16cm]{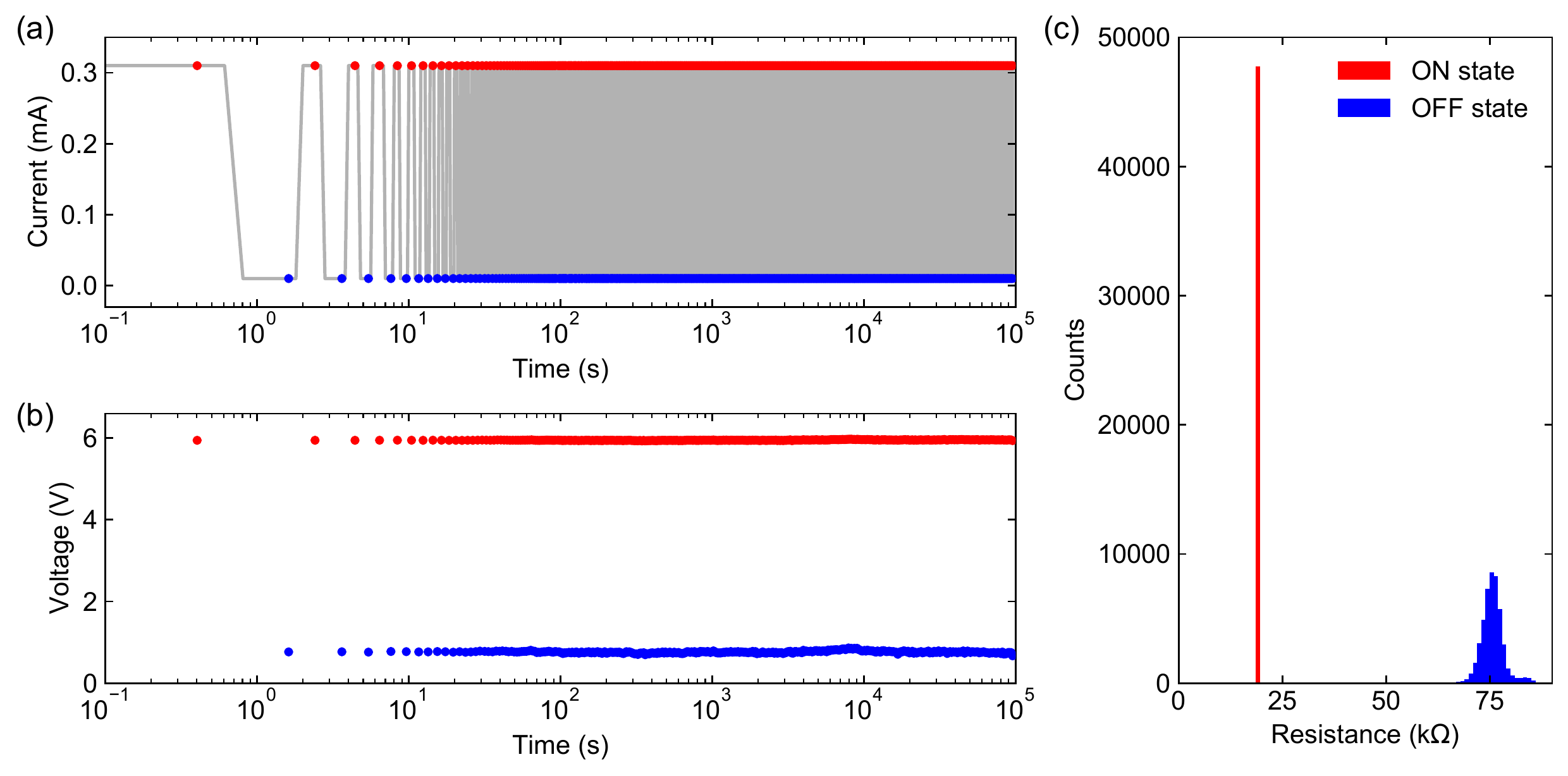}
\end{center}
\caption{
\rev{Current switching cycles on \crto\ $x=0.5\%$ (sample \#2). (a) The current (gray line) was switched back and forth between the ON state (0.31 mA) and the OFF state (0.01 mA) every second for over $10^5$ sec., and (b) the voltage across the sample was recorded at each cycle (red and blue circles). (c) Histogram of the two-probe resistance in the ON and OFF states. The resistance spread in the OFF state is due to the temperature fluctuations}   
\label{fig:switching-hystogram}
}
\end{figure*}
%%%%%%%%%%%%%%%%%%%%%%%%%%%%%%%%%%%%%%%%%%%%%%%%%%%%
\begin{figure*}
\begin{center}
\includegraphics[width=12cm]{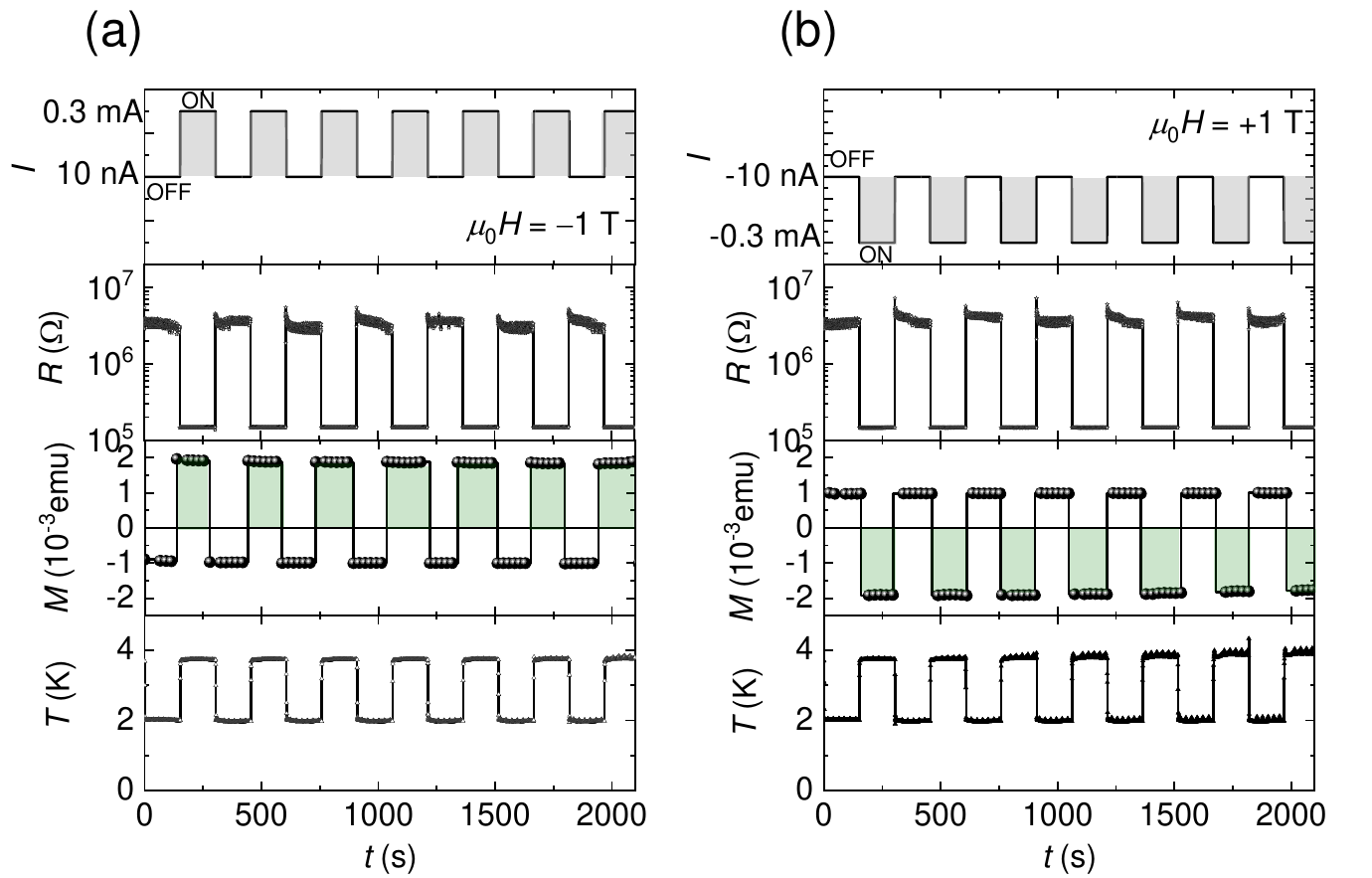}
\end{center}
\caption{
\rev{Simultaneous resistance and magnetization control by DC current in \crto\ $x=0.5\%$ sample \#1}. (a) In a negative field of $-1$~T (current changed between 10~nA and 300~$\mu$A), the magnetization is positive in the ON state, anti-parallel to the external field as it happens for positive field of 1 T. (b) For negative current between $-10$~nA and $-300$~$\mu$A (at +1~T), the magnetization in the ON state is negative, indicating that the magnetization direction does not depend on the current direction. The sample temperature shows a slight increase of 2 K in the ON state due to Joule heating, but its effect is negligible in the large observed changes of $R$ and $M$.   
\label{fig:switching-details}
}
\end{figure*}
\clearpage

\end{document}